# Deep learning and traditional-based CAD schemes for the pulmonary embolism diagnosis: A survey

**Seyed Hesamoddin Hosseini** [a], **Amir Hossein Taherinia** [a,*], **Mahdi Saadatmand** [b]

[a] *Department of Computer Engineering, Ferdowsi University of Mashhad, Mashhad, Iran*

[b] *Rayan Center for Neuroscience and Behavior & Medical Imaging Lab, Department of Electrical Engineering, Faculty of Engineering Ferdowsi University of Mashhad, Mashhad, Iran*

*Email: hesamoddin.hosseini@mail.um.ac.ir, taherinia@um.ac.ir, saadatmand@um.ac.ir*

*** Corresponding Author**

## Abstract

Nowadays, pulmonary Computed Tomography Angiography (CTA) is the main tool for detecting Pulmonary Embolism (PE). However, manual interpretation of CTA volume requires a radiologist, which is time-consuming and error-prone due to the specific conditions of lung tissue, large volume of data, lack of experience, and eye fatigue. Therefore, Computer-Aided Design (CAD) systems are used as a second opinion for the diagnosis of PE. The purpose of this article is to review, evaluate, and compare the performance of deep learning and traditional-based CAD system for diagnosis PE and to help physicians and researchers in this field. In this study, all articles available in databases such as IEEE, ScienceDirect, Wiley, Springer, Nature, and Wolters Kluwer in the field of PE diagnosis were examined using traditional and deep learning methods. From 2002 to 2023, 23 papers were studied to extract the articles with the considered limitations. Each paper presents an automatic PE detection system that we evaluate using criteria such as sensitivity, False Positives (FP), and the number of datasets. This research work includes recent studies, state-of-the-art research works, and a more comprehensive overview compared to previously published review articles in this research area.

## Introduction

Pulmonary embolism (PE) is one of the leading causes of death in the world, and early detection can save many lives [1]. PE is a clot in the blood vessels caused by a sudden occlusion of the pulmonary artery. These blood clots usually form in the lower veins of the pelvis and travel via the blood flow, passing through the heart, to the blood arteries of the lung tissue, where they settle. This phenomenon causes the closing of the pulmonary artery and thus reduces the ability to breathe [2]. Sudden death occurs when a large clot blocks more than 50% of blood flow. Therefore, PE is a common disorder with high mortality complications, which requires early and accurate diagnosis. Contrast-enhanced X-ray Computed Tomography (CT) images are widely used in the diagnosis of PE because these images have a low risk and a high accuracy rate, and also show good representation of the lesion in the blood vessels. The contrast agent that is injected into the patient dissolves in the blood and increases the contrast of the vessels, and the vessels appear as a bright area in Computed Tomography Angiography (CTA) images, but this substance does not dissolve with embolism. Therefore, it appears as a dark area in CTA images [3]. Manual identification of dark spots related to emboli by a radiologist is a difficult and time-consuming task, so that the same masses may not be identified by several radiologists with the same data set. Hence, Computer-Aided Diagnosis (CAD) systems have gained great importance to support radiologists and improve their performance in the challenge of CTA images [4]. In recent years, a lot of research has been done in the field of designing automatic PE detection systems. The CAD system consists of three parts as shown in the figure 1. The CAD models look for tissue changes and abnormalities like radiologists. First, they reduce the search space by an operational set (lung extraction); then, considering the assumptions, they identify a set of candidate points for embolism (Candidate Detection). In the next step, according to the different features of embolism in the lung anatomy, they try to extract different features in different spatial conditions (Feature Extraction) and in the last step, by reducing the positive error, PE areas are separated from other background areas (Classification) [5].

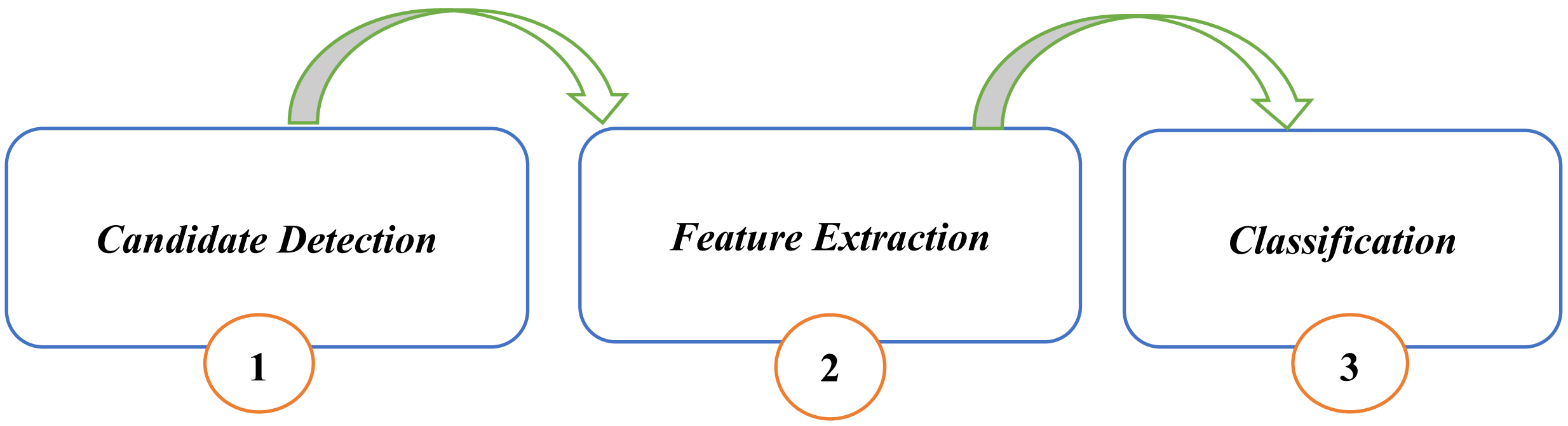

Fig. 1. CAD system design step

Handcrafted feature extraction engineering, such as texture and shape analysis, is the foundation of traditional CAD systems. Hand-crafted characteristics are derived from direct visual experience and may be ineffective due to their abstraction and construction from low-quality medical image samples. Furthermore, they lack uniformity, standardization, and universality, and they take an inordinate amount of time.

Deep learning approaches, particularly Convolutional Neural Networks (CNNs), have recently received a lot of interest from researchers due of their accuracy and time savings when compared to traditional CAD systems. The CNN approach has discovered a solution to a variety of learning problems, including the extraction of relevant features, object recognition, accurate detection, and so on. It has sufficiently demonstrated its performance in a variety of disciplines, including medical image analysis [6].

In the proposed taxonomy, various models are offered that exists different False Positive (FP) and sensitivity for diagnosing PE. The CAD system, as shown in Figure 1, consists of candidate detection, feature extraction, and classification. In this research, we categorize the presented articles in to two main sections that includes deep learning and traditional methods for PE diagnosis in figure 2.

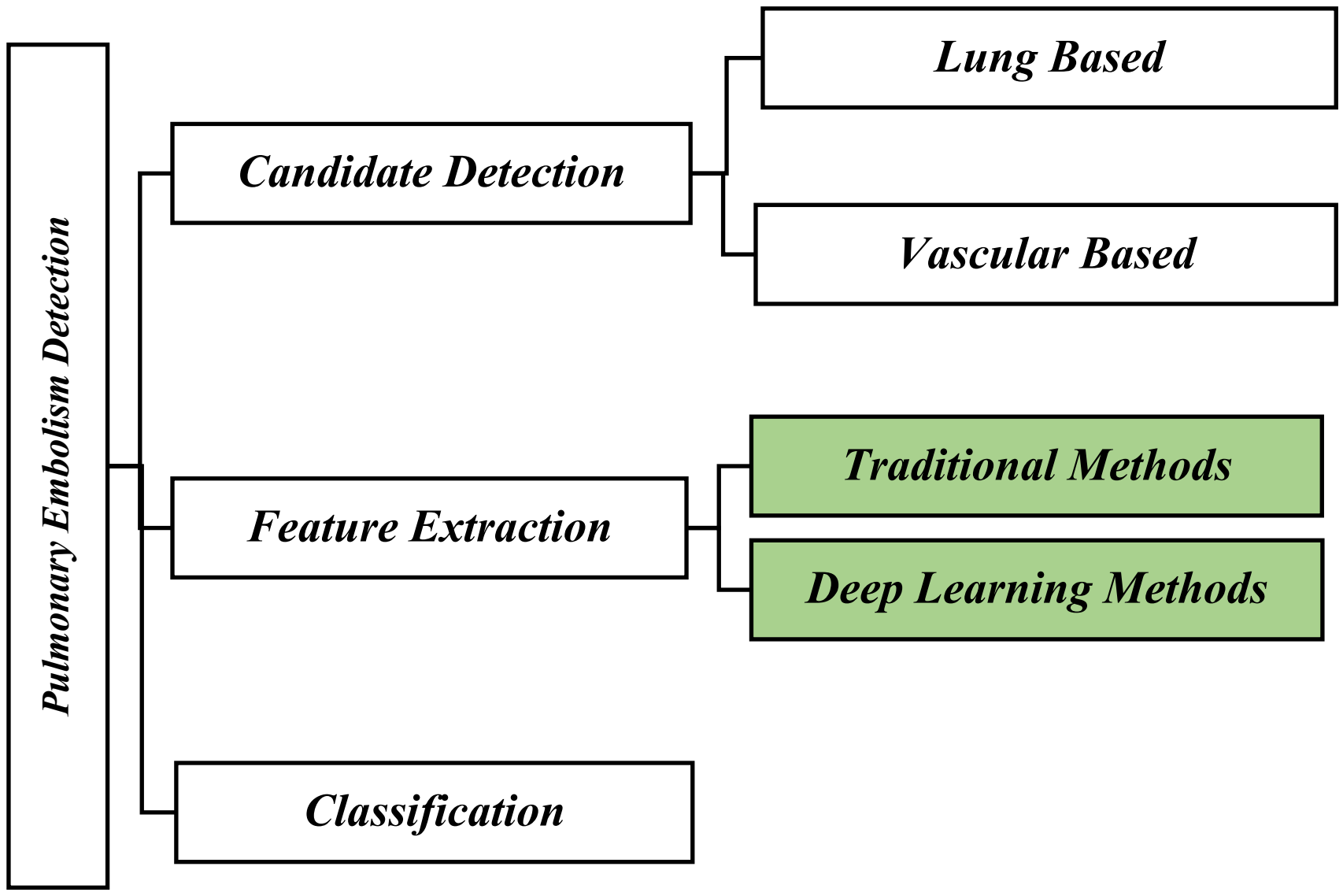

Fig. 2. The taxonomy of PE detection

This research work provides a comprehensive review on previous state-of-the-art research works based on deep learning and image processing techniques in diagnosing PE. We first extract the existing articles in this field with a suitable search in Table 1. Then, there will be a discussion about how to analyze and categorize the articles. Conclusions about PE detection systems cannot be specified with a few quantitative criteria because different parameters are involved in the results; Therefore, in this paper, in addition to the quantitative evaluation of the results, the qualitative and analytical evaluation of the results will also be done and these results will be discussed at the end. The sensitivity of the results of the PE detection systems depends on the image database. Our goal in this article is to review automatic PE detection systems and evaluate their performance.

The rest of the article is organized as follows. Section 2 describes research methodology. Section 3 literature review and Section 4 analysis based on literature review. Finally, Section 5 and 6 presents the results and conclusion, respectively. section 7 about limitation and future work.

## 2. Research Methodology

This research is a review study in the field of PE diagnosis using deep learning and traditional-based CAD systems. In order to select suitable papers in this field, first a comprehensive and complete search was conducted. The terms used in Table 1 were used to search for the obtained article, and this phrase was the same in all databases such as Springer, Nature, ScienceDirect - Elsevier, IEEE, Wolters Kluwer, and Wiley from 2002 to 2023. Physicians use CTA images to diagnose embolism, so according to the choice of the keyword pulmonary embolism diagnosis, CTA is automatically considered and there is no need to specify the type of images.

Inclusion Criteria Search (ICS) and Exclusion Criteria Search (ECS) aim to identify studies related to more details. We reviewed 23 English articles in journals from 2002 to 2023. Table 2 explains the ICS and ECS papers.

Table 1. Searched terms

| | |
|---|---|
| S1 | "Document Title": ***(Pulmonary Embolism***) AND "Title": (***Deep Learning***) |
| S2 | "Document Title": ***(Pulmonary Embolism***) AND "Abstract": (***Deep Learning***) |
| S3 | "Document Title": ***(Pulmonary Embolism***) AND "Title": (**Image Processing**) |
| S4 | "Document Title": ***(Pulmonary Embolism***) AND "Abstract": (**Image Processing**) |

Table 2. ICS and ECS

| **Inclusion Criteria** |
|---|
| *Studies on different methods of early-stage PE*<br>*Studies have been published from 2002 to 2023* |
| **Exclusion Criteria** |
| *Studies that focus on books and technical reports*<br>*Studies that are not in English*<br>*Studies in which claims focus rather than evidence* |

From 2002 to 2023, 23 articles on the diagnosis PE in the listed databases were selected and thoroughly studied. One such article in 2021, Soffer et al. [7] presented a systematic review article. Figure 3 illustrates the number of selected articles on PE diagnosis by year and Figure 4 demonstrates the number of journal articles in databases.

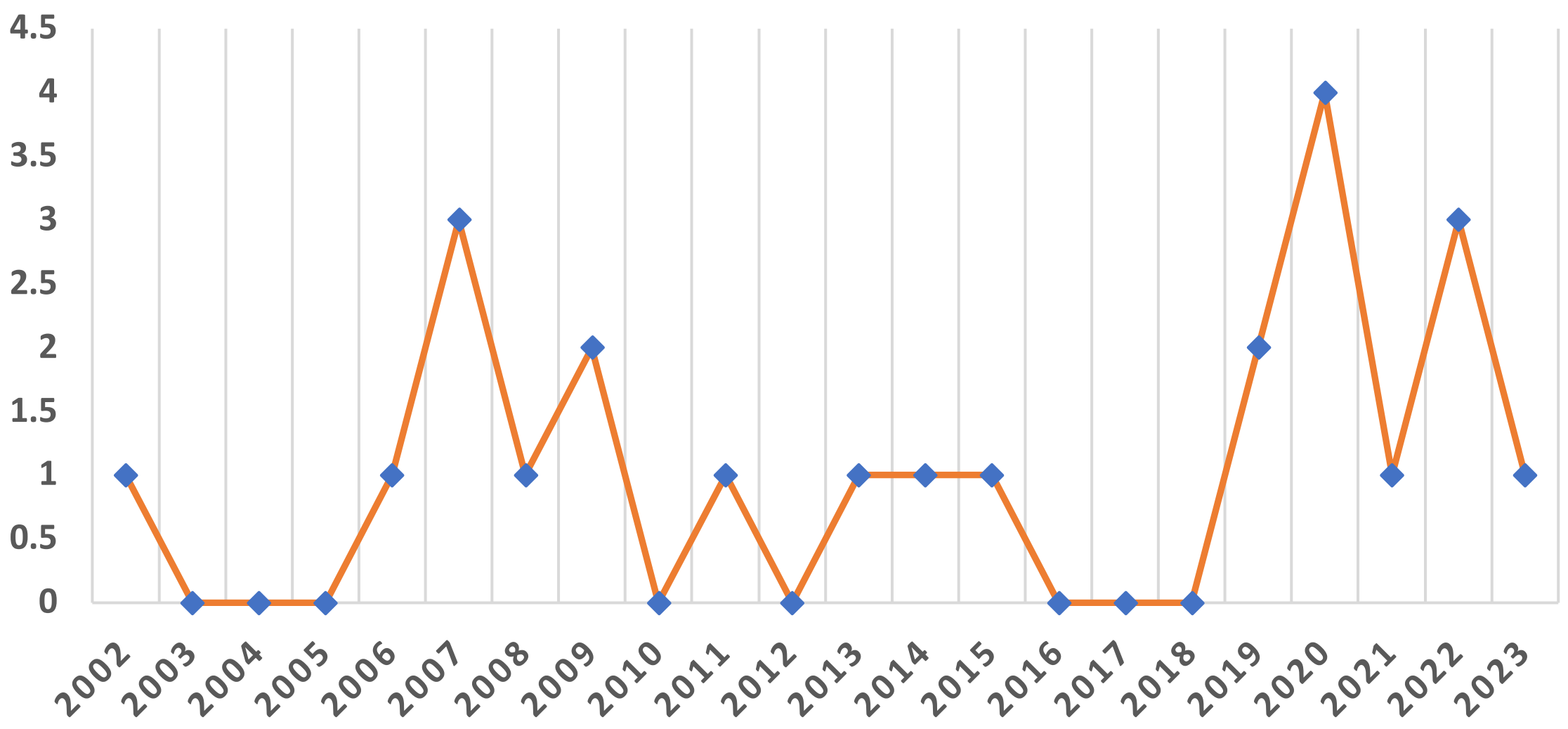

Fig. 3. The number of articles on PE diagnosis by year

As shown in Figure 3, 2020 is the highest study in this field. This shows that this topic is increasing every year, which we think is of particular importance.

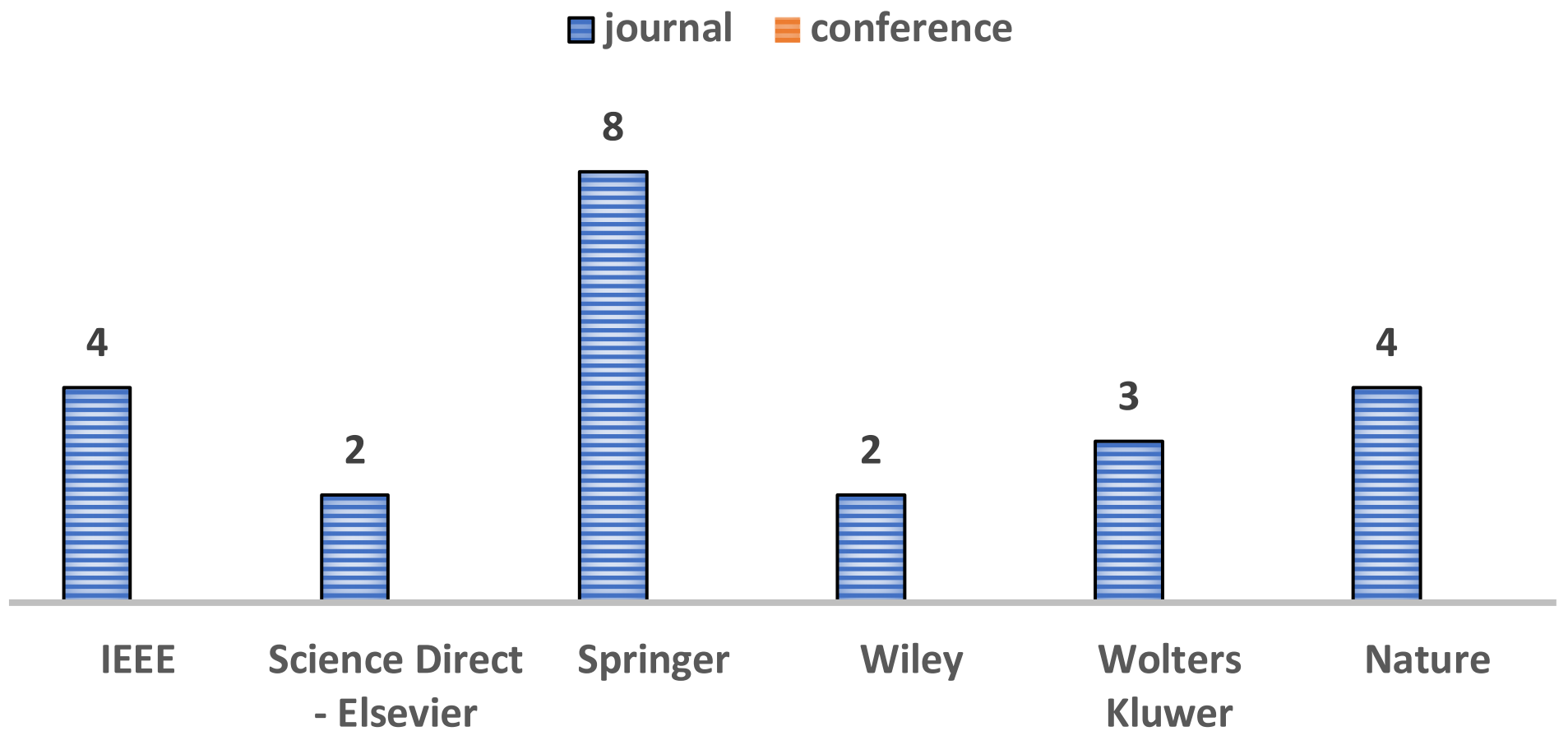

Fig. 4. The number of selected journal articles on PE diagnosis

As can be seen in Figure 4, Springer has the highest number of published articles on PE diagnosis. Wiley and Science Direct - Elsevier also have the fewest articles published in this field.

Next, we tried to categorize the papers. A suitable criterion for classification can be the type of benchmark dataset. Considering that we know that PE if it occurs in the secondary branches of the vessels, it is much more difficult to diagnose it than in the main branches. Therefore, the selection of test samples has a significant impact on the accuracy of articles. If the databases of each of the articles were the same or at least specified the number of emboli in the sub-branches, the task of classifying the articles would be done easily. In reality this was not the case and each paper used a local dataset. Another criterion could be the number of data or the number of emboli, but considering the variety of emboli in the main or secondary branches of the vessels, this criterion was not suitable. Therefore, we did not limit ourselves to the type or number of databases to categorize the articles and examined each of the systems designed to detect PE separately.

Feature extraction methods help a lot to describe emboli. In this research, to classify PE diagnosis papers, they are divided into two categories: Hand-crafted feature extraction (traditional-based CAD) and automatic feature extraction (deep learning-based CAD). As shown in Figure 5, 11 papers used traditional-based CAD and 12 papers used deep learning-based CAD to detect PE. We have categorized each of these methods below.

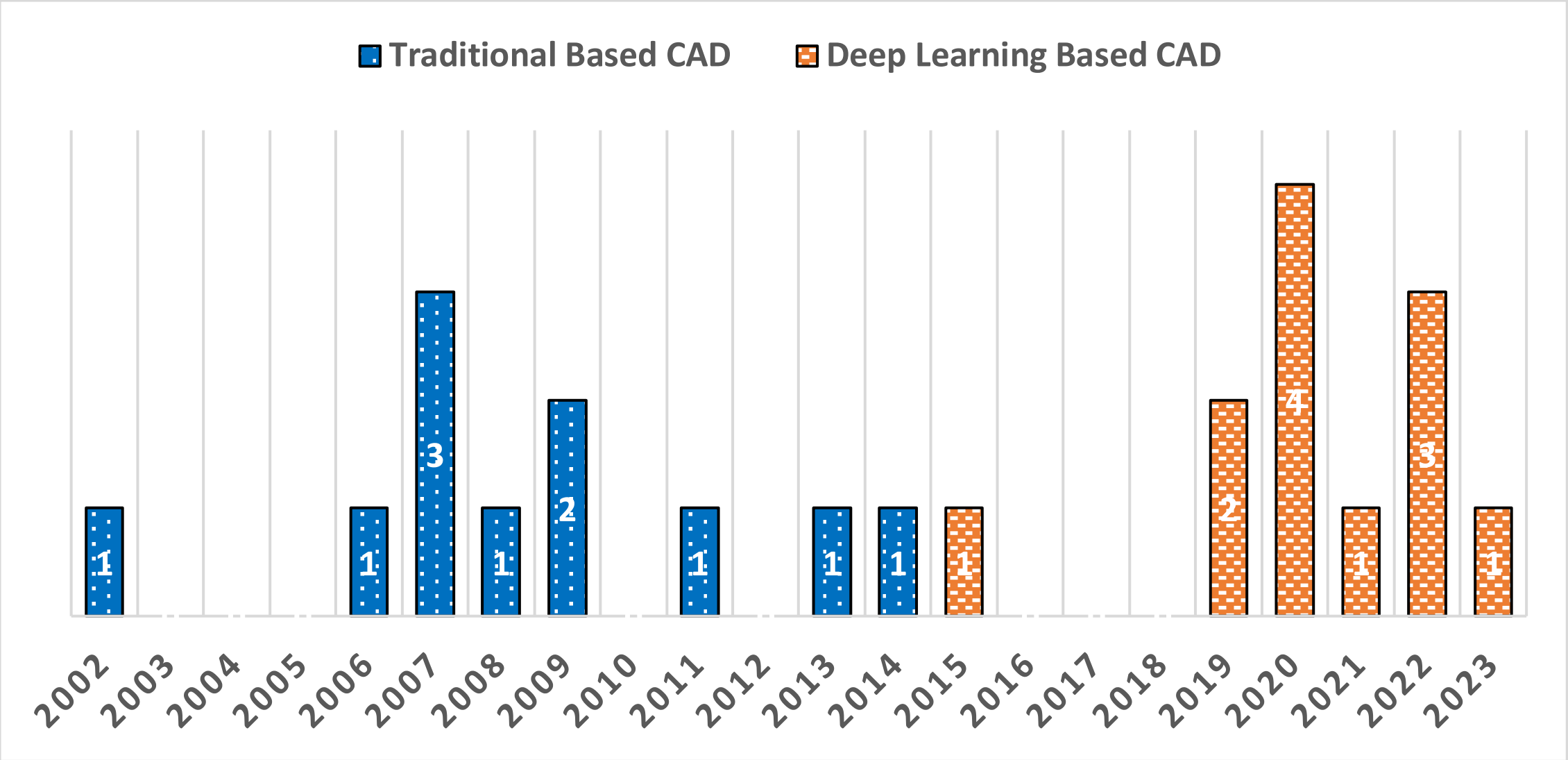

Fig. 5. Deep learning and traditional-based CAD articles

As shown in Figure 5, The initial work in the diagnosis of PE was based on traditional methods [8,9]. However, modest success has been achieved with limited clinical application. These techniques have only been tested on small groups. The advent of deep learning has increased interest in automated emboli detection in radiology since 2015. Applications of deep learning have already shown promise in medical imaging, including nodule detection in chest X-ray images [10], brain hemorrhage detection in CT scans [11], and tumor detection in Magnetic Resonance Imaging (MRI) [12].

## 3. CAD system design

In most of the works selected in this literature review, CAD system design consists of three main stages: candidate detection, feature extraction, and classification. Overall, feature extraction is done both manually and automatically. If feature extraction is done manually, it is called traditional-based CAD system, otherwise it is called deep learning-based CAD system [13]. First, the articles whose features were extracted manually (traditional-based CAD system) and published between 2002 to 2014 are reviewed. Then, the articles whose features were automatically extracted (deep learning-based CAD system) and published between 2015 to 2023 are reviewed.

### 3.1 Traditional-based CAD system

In the years 2002 to 2014, each of the methods have somehow tried to correctly express the behavior of embolism in pulmonary arteries. We classify these features into four categories: intensity-based, shape-based, position-based, and boundary-based.

The first characteristic that can be considered for the intensity-based category is the intensity value of the Hounsfield CT [14,15,16,17]. This value is different for each data set and should be determined accordingly. Several morphological operations such as bottom-hat transform [18] have also been used as a suitable feature. This transformation finds small dark spots around bright spots. First, Masutani et al. [14], then Bouma et al. [15] and Zhou et al. [16] respectively used the eigenvalues of the Hessian matrix. If we have a dark area in the middle of a light background, the eigenvalue of the Hessian matrix must be positive. Other methods such as pixel fluctuation rate, area resolution, the maximum value of the histogram of a local area, etc. have also been used, which are shown in Table 3-A.

The degree of embolus curvature is one of the important features used in shape-based methods. The degree of curvature [19] shows the shape of a plane that passes through a point. Under normal conditions, the vein has a tubular shape that resembles a bump. The vein is saddle-shaped in its branches, but the lighting shape around the embolus is shaped like a pit. This criterion indicates whether the candidate point is inside or outside the vein. The eigenvalues of the Hessian matrix have also been used to calculate the curvature degree characteristic [14,15,16]. This characteristic does not always give the correct answer because the emboli may not have a linear shape in the branches. In Table 3-B, the different methods of feature extraction based on the shape are fully described.

Bouma et al. [15] first used position-based features for vessel extraction. This feature is introduced as the distance to the main texture. The distance between the embolus and the parenchyma is greater than the dark areas around the vessels (Table 3-C). In Table 3-D, Park et al. [20] proposed a feature based on the boundary of the emboli region. Different hand-crafted feature extraction methods can be seen in Table 3 and traditional based CAD systems for identifying PE are presented in Table 4.

Table 3. Different hand-crafted feature extraction

| | |
|---|---|
| A) **Intensity-based** | Pixel size of the embolism candidate area (2D or 3D) [15] |
| | Eigenvalues of the Hessian matrix [15] |
| | Bottom-hat transform [15] |
| | Maximum histogram value [16] |
| | Morphological characteristic between the embolic candidate area, surrounding vessels and non-vascular background [16] |
| | Contrast the area with the surrounding vessels and with the non-vascular background [16] |
| | Voxel local contrast [14] |
| B) **Shape-based** | Lumen curvature [15] |
| | The degree of roundness [15] |
| | Length and mean radial area [20] |
| | The distance from the center of gravity to the Hounsfield maximum [20] |
| | Width and height of the area [20] |
| | The degree of curvature [14] |
| | Effective length [14] |
| | Vascular radius [14] |
| C) **Position -based** | Distance to the main texture [15] |
| | Connection with other vessels [15] |
| D) **Boundary -based** | The standard deviation of the gradient of the border pixels [20] |
| | Kurtosis gradient of border pixels [20] |

Table 4. An overview of traditional-based CAD systems for identifying PE

| # | Ref | Year | Method |
|---|---|---|---|
| 1 | Masutani et al. [14] | 2002 | Based on vessel extraction using thresholding and region growing, Feature extraction |
| 2 | Digumarthy et al. [21] | 2006 | Based on lung extraction, Feature extraction |
| 3 | Liang et al. [22] | 2007 | Based on lung extraction, Using the Tobogganing algorithm, Feature extraction, Feature reduction using support vector machines |
| 4 | Buhmann et al. [23] | 2007 | Designing a computer-aided recognition system, Including Tobogganing algorithm, Feature extraction |
| 5 | Maizlin et al. [24] | 2007 | Vessel extraction with user interface |
| 6 | Das et al. [25] | 2008 | Designing a computer-aided recognition system, Including Tobogganing algorithm, Feature extraction and Positive error reduction filters |
| 7 | Bouma et al. [15] | 2009 | Based on vessel extraction using tracking, Position-based method, Using tree classification |
| 8 | Zhou et al. [16] | 2009 | Based on vessel extraction, Feature extraction, Linear discriminant classifier |
| 9 | Park et al. [20] | 2011 | Based on lung extraction, Using the Tobogganing algorithm, Feature extraction, Reduced specificity with genetics, Classification with Artificial Neural Network (ANN) |
| 10 | Wittenberg et al. [26] | 2013 | Lung extraction, Analysis and investigation of complete occlusion vessels |
| 11 | Ozkan et al. [27] | 2014 | Based on vessel extraction using lung anatomy information, Feature reduction to reduce false positives |

## 3.2 Deep learning-based CAD system

Deep learning models can diagnose PE in CTA images with satisfactory sensitivity (acceptable number of FP [7]). In 2015, Tajbakhsh et al. [28] were the first to use CNN to diagnose PE. To successfully use a CNN to detect emboli and eliminate FP in 3D images is to develop a "right" image representation for the object. Toward this end, they proposed a vessel-aligned multi-planar image representation of emboli to create a two-channel image representation for each embolism candidate. Then, the resulting two-channel patches have been fed to a CNN to classify the candidates into PE or non-PE categories. A vessel-aligned multi-planar image representation of emboli that offers three advantages: (1) efficiency and compactness because it briefly summarizes the 3D contextual information around an embolus in only two channels; (2) consistency because it automatically aligns the embolus in the two-channel images according to the orientation of the affected vessel; and (3) expandability because it naturally supports data augmentation for training a CNN. Experimental results show that this CAD system has a sensitivity of 83% at 2 FP per volume of 121 CT pulmonary angiography (CTPA) datasets with a total of 326 emboli.

Tajbakhsh et al. [29] developed their previous work and proposed a novel vessel-oriented image representation (VOIR). The proposed image representation has four advantages, which are: (1) efficient and compact because it briefly summarizes the 3D contextual information around an embolus in only three channels; (2) consistent because it automatically aligns the embolus in the three-channel images according to the orientation of the affected vessel; (3) expandable because it naturally supports data augmentation for training a CNN; and (4) multi-view visualization is used in their PE visualization system. They showed that different CNN architectures trained using a novel VOIR can perform significantly better than their counterparts trained using standard and 2.5D image representations.

Huang et al. [30] developed a PENet deep learning model for automatic emboli detection in volumetric CTA as an end-to-end solution for this purpose. PENet is a 77-layer 3D CNN pre-trained on the Kinetics-600 dataset [31] and fine-tuned on a CTPA set collected from an academic institution. PENet is built using four architectural units, which are: (1) the PENet unit is used to process 3D input data; (2) Squeeze-and-Excitation (SE) block is used to model interdependencies between input channels and adaptively recalibrate channel properties; (3) the PENet bottleneck is built using two PENet units and SE block; and (4) the PENet encoder includes sequence of multiple PENet bottlenecks, ranging from three to six [32]. The advantages of this work are: (1) development 3D CNN model (open-source) for PE detection on patient data from two hospital systems (Stanford and intermountain); (2) enable others to reproduce and allow for further innovation this method by an outcome-labelled CTPA dataset; (3) an end-to-end model that can acquire CTPA volumetric imaging studies without the need pre-processing or feature engineering; (4) for the timely diagnosis of this important disease, including in places where radiology expertise is limited.

Recent advances in deep learning have led to the resurgence of medical imaging and Electronic Medical Record (EMR) models for various applications, including clinical decision support, clinical prediction, and more. Huang et al. [33] have developed and compared a multimodal fusion deep learning model that uses information from CT and EMR images to automatically detect pulmonary embolism. The use of clinical and imaging data using a variety of fusion methods can not only lead to a contextually relevant model that reduces the misdiagnosis and delay of embolism. Instead, it informs future work by examining data selection and optimal fusion strategies. The imaging model is the same as PENet described in previous work and the EMR model is a simple feedforward neural network that uses a combination of all EMR features as input (except CT imaging features). The processed data of imaging model and EMR model were also used for fusion models. They have implemented different fusion architectures, including an early fusion model, four different types of late fusion models, and two Joint fusion models that take advantage of patient CT scan and EMR information. The best performance is related to the late fusion model implemented using 3D neural network and ElasticNet which achieved an AUROC of 0.962.

To calculate the emboli volume, the clot areas must be extracted from the image. Therefore, Liu et al. [34] developed a deep learning model based on the U-Net framework to automatically segment clot regions from CTA images and also calculate the clot burden of Acute Pulmonary Embolism (APE). Their proposed method is performed in two steps: first, dividing the clot and then calculating the volume of the clot. To detect clots, different probability thresholds have been determined, and if the areas whose probability values are higher than the threshold, they are considered clots. The U-Net model can diagnose clot with 94.6% and 76.5% sensitivity and specificity respectively.

Weikert et al. [35] have proposed an algorithm based on a trained Deep Convolutional Neural Network (DCNN) and evaluated it on 2800 data, which is a large dataset, for the automatic detection of pulmonary embolism in CTA. The proposed algorithm consists of two steps: (1) a region proposal stage; and (2) a FP reduction stage. In the first stage, it is a 3D DCNN, whose architecture is based on the Resnet architecture [36], which consists of multi-layer convolution blocks with skip connections between them, followed by a pooling layer. This network is trained on segmented scans and produces a 3D segmentation map. The segmentation map, region proposals are generated and transferred as input to the second stage of the algorithm. The second stage classifies each region as positive or negative based on the features of the last layer of the first stage and traditional image processing methods. Their artificial intelligence algorithm has high accuracy for detecting embolism in CTA.

Yang et al. [37] presented a two-stage CNN for automatic detection of emboli in CTA with 129 data (a total of 269 embolisms). For the first time, all steps in this method, including emboli candidate proposal, generation of vessel-aligned image representation, and FP removal, have been implemented in a two-stage CNN for a highly accurate emboli detector. Their PE diagnosis network consists of a cascade two stages: (1) The first stage is a proposed 3D sub-network based on a Fully Convolutional Neural Network (FCN), which aims at high sensitivity and logical FP; (2) The second stage is a FP sub-grid based on vessel alignment, which aims to remove as many FPs as possible through a classification while maintaining high sensitivity.

Deep Venous Thrombosis (DVT) causes embolism, which is considered a serious cardiovascular problem that can lead to death if not treated properly. Lynch et al. [38] Proposed a CNN model called Pulmonary Embolism detection using Deep Neural Network (PE-DeepNet) for effective prediction and classification of PE. First, 512*512 input images were resized to 640*640 and a pre-processing was performed to improve the location of the lung portion in the entire image and improve the computation. Then the images are normalized in a batch and fed to the proposed neural network. Next, the feature extraction process is carried out and given to the binary classifier. The architecture of their proposed CNN model consists of two 2D convolutional layers with activation function as Rectified Linear Unit (ReLU), a global max-pooling layer, PE-DeepNet is an additional block (two 2D convolutional layers with activation function as ReLU, batch normalization, and a global max-pooling layer), one dropout layer, one flatten layer, three dense layers, and a sigmoid classifier. Their proposed model can diagnosis PE with 94.2% accuracy of the standard Radiological Society of North America (RSNA) dataset in association with the Society of Thoracic Radiology (STR) [39].

Huhtanen et al. [40] have developed a deep neural network model consisting of two parts. A CNN architecture called InceptionResNet V2 and a long-short term memory (LSTM) network, which sequentially processes the slice predictions generated by the CNN section. Their proposed method was to first rescale the 512*512 images to 386*386, which is the required input size of the CNN model. The main advantage of this approach is that 2D convolutional neural network models are significantly simpler and require less data than 3D convolutional neural network models. Evaluation experiments show that the proposed model can detect PE with 93.5% sensitivity.

Ma et al. [41] have proposed a two-phase multitasking learning method that can detect the presence of embolism and its characteristics such as location of PE (left, right, or central), PE condition (acute or chronic). The first phase uses a 3D CNN for feature extraction and a temporal convolutional network (TCN), with attention mechanisms in the second phase to perform sequential learning. TCN consists of two dilated convolution layers, two normalization layers, two ReLU activation, and two dropout layers. Attention mechanisms are used to assign weights to features in a sequence, where higher weights indicate a higher probability of embolism. Their proposed model can diagnosis PE with 86% sensitivity and 85% specificity of the RSNA–STR pulmonary embolism CT dataset

The methods studied by others in the field of automatic embolism detection are almost all two-step, which involve searching for candidates and then classifying the candidates. Also, the pre-processing process of some of these studies is tedious, which increases the calculation time to some extent. Therefore, Xu et al. [42] have proposed a deep learning model of YOLO version 4 for the automatic detection of embolism. Yolo Editions was born as a one-step target detection algorithm to provide a fast and accurate target detector and has been well received since its release in 2016 [43]. Therefore, they used YOLO version 4 with 307 included databases (142 patients from Tianjin database, 133 patients from Linyi database and 32 patients from FUMPE database) to diagnose embolism. According to the presented results, this method works well in the automatic diagnosis of embolism and provides the results in a short period of time. In other words, the model-assisted diagnosis results are available in real time at the end of the patient examination, providing a reference for the radiologist's diagnosis. YOLO version 4 is a two-dimensional deep convolutional network. Using pre-processed CTA images as input, the neural network can directly detect emboli. Their embolic detection network consists of four modules. The input module is used to merge data and image labels, data enhancement, etc. The CSP module is used for feature extraction and the SPP module is used for merging the extracted features. Finally, the prediction module aggregates the features to obtain the final prediction results. Their proposed model can detect PE with 95% sensitivity.

## 4. Analysis of the reviewed of the deep learning and traditional-based CAD system

Two different analyses were performed to evaluate PE diagnosis based on deep learning and traditional methods:

The first analysis of classification is based on a literature review (Fig.6).
The second analysis the number of patients, PE-Clots, FP, sensitivity, and publicly available datasets are based on literature review (Table 5).

### 4.1 Summary of taxonomy based on literature review

According to the taxonomy presented in the introduction (Fig.2). After reviewing the articles, we explained the relationship of each of these articles with the classification presented in Fig.6. The reviewed articles are classified according to the deep learning and traditional methods shown in Fig. 6.

### 4.2 Summary of primary studies based on literature review

We thoroughly review and analyze early PE studies based on deep learning and traditional methods. As can be seen in Table 5, an overview of the comparison of traditional methods and deep learning methods is presented.

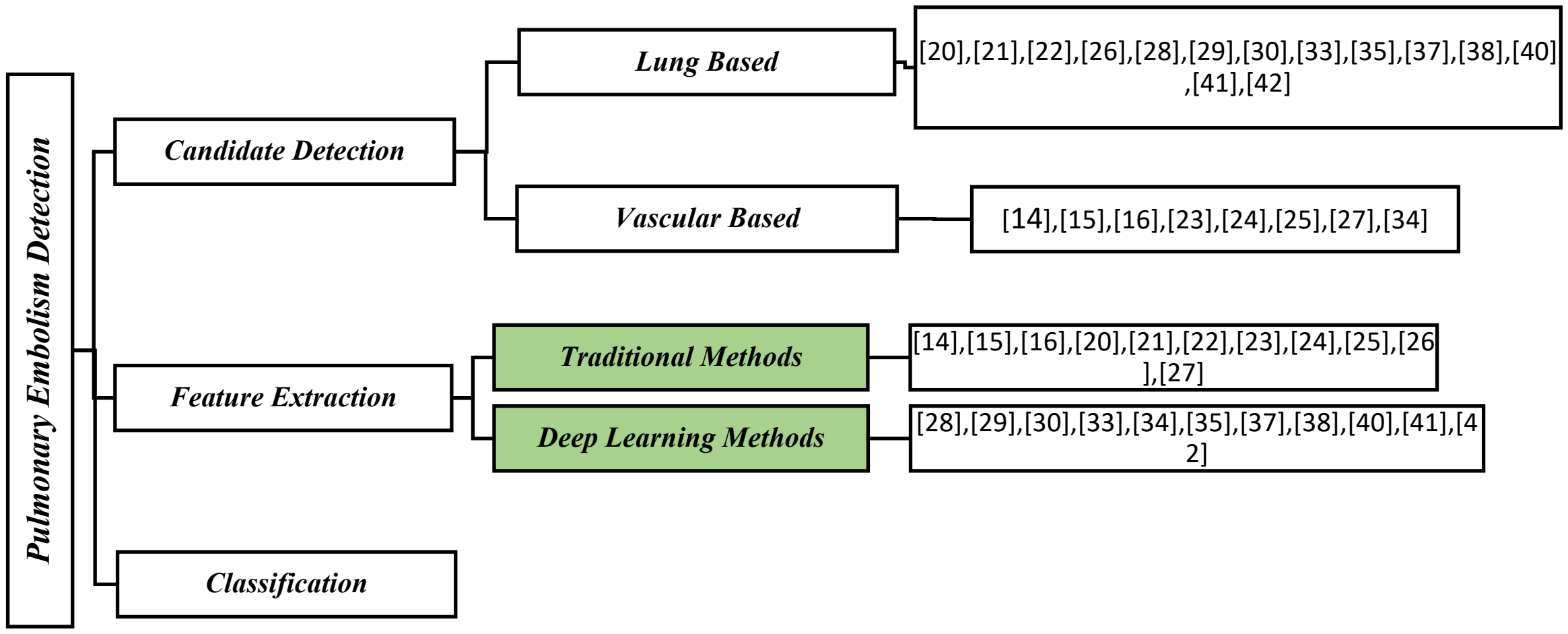

Fig. 6. The taxonomy of PE detection based on reviewed papers

Table 5. Performance of deep learning and traditional-based CAD systems for PE diagnosis on literature review

| | Ref | Year | Patient | PE-Clots | Sensitivity | FP | Publicly available datasets | Available Scoring |
|---|---|---|---|---|---|---|---|---|
| Traditional Methods | Masutani et al. [14] | 2002 | 19 | 21 | 100.0/85.0 | 7.7/2.6 | × | × |
| | Digumarthy et al. [21] | 2006 | 39 | 270 | 92.0% | 2.8 | × | × |
| | Liang et al [22] | 2007 | 177 | 872 | 80.0% | 4.0% | × | × |
| | Buhmann et al. [23] | 2007 | 40 | 212 | 70.0% | 9.0% | × | × |
| | Maizlin et al. [24] | 2007 | 8 | 45 | 58.0% | 6.4 | × | × |
| | Das et al. [25] | 2008 | 33 | 288 | 88.0% | 4.0 | × | × |
| | Bouma et al [15] | 2009 | 19 | 116 | 58.0%<br>63.0%<br>73.0% | 4.0<br>4.9<br>15.0 | × | × |
| | Zhou et al [16] | 2009 | 128 | 1395 | 58.0% | 10.5 | × | × |
| | Park et al [20] | 2011 | 20 | 648 | 63.2% | 18.4 | × | × |
| | Wittenberg et al. [26] | 2013 | 38 | 119 | 95.0% | 3.8 | × | × |
| | Ozkan et al [27] | 2014 | 33 | 450 | 61%<br>95% | 8.2<br>14.4 | × | × |
| **Deep Learning Methods** | Tajbakhsh et al 0 | 2015 | 121 | 326 | 83.4% | 2.0 | × | × |
| | Tajbakhsh et al [29] | 2019 | 121 | 326 | 83% | 2.0 | × | × |
| | Huang et al [30] | 2020 | 1773 | 1797 | 75% | - | The datasets generated are not publicly available but are available from the corresponding author on responsible request | × |
| | Huang et al [33] | 2020 | 1794 | 2563 | 87.3% | - | The datasets generated are not publicly available but are available from the corresponding author on responsible request | × |
| | Liu et al [34] | 2020 | 878 | 646 | 94.6% | - | × | Q-Score, Mastora Score |
| | Weikert et al [35] | 2019 | 1465 | 234 | 92.7% | - | × | × |
| | Yang et al [37] | 2019 | 129 | 269 | 75.4% | 2.0 | × | × |
| | Lynch et al [38] | 2022 | 9446 | - | 93.8% | - | × | × |
| | Huhtanen et al [40] | 2022 | 600 | - | 93.5% | - | × | × |
| | Ma et al [41] | 2022 | 7279 | - | 86% | - | ✓ | × |
| | Xu et al [42] | 2023 | 307 | - | 95% | - | × | Q-Score |

## 5. Results

The success of embolism detection systems cannot be measured by one factor; Therefore, in this paper, we tried to examine the designed systems from different aspects and evaluate them. As show in Table 5, deep learning methods have had a significant impact on embolism detection in 2015 because:

- Radiologists provide a sensitivity of 0.67 to 0.87 with a specificity of 0.89 to 0.9931 for PE detection, while deep learning models that take an automated approach to PE detection have a sensitivity of 0.88 and a specificity of 0.86 [44.45].
- Accurate and rapid diagnosis of embolism is essential to improve prevention.
- A high number of FP cases causes fatigue [46]. Finally, an Artificial Intelligence (AI) tool has the potential to reduce the time to detect PE.
- It is important to evaluate whether an automated PE detection system can improve radiologist performance and ultimately lead to better clinical outcomes.

## 6. Conclusion

In this paper, twenty-three articles were reviewed and different methods for PE detection with deep learning and traditional methods were presented, resulting in different levels of sensitivity and FP. This article examines deep learning and traditional-based CAD system to PE detection. Most publications in this area are attributed to Springer and are the most important study in this area in 2020. The current article has limitations because the search was only performed on the title of the article. If you expand this search, you can access other articles. This study focused only on English-language publications and no search for non-English-language articles was conducted.

## 7. Limitation and future work

To draw inferences from existing algorithms, some constraints must be mentioned:

1- Each of the image processing algorithms are numerically and statistically comparable if they are applied to single images and the conditions are the same, but it is not the same in the available articles of the image bank.
2- As stated in the CAD system design section of the article, the authors of the articles looked at solving the embolism problem in two categories: deep learning and traditional-based CAD system. This completely different view of the problem makes statistical comparisons difficult.
3- In the image bank, the presence or absence of various lung diseases has a significant impact on the process of diagnosing embolism. If a sample has a disease other than embolism, it becomes difficult to find embolism, so this criterion is also checked in the Analysis of the reviewed of the deep learning and traditional-based CAD system section.

Despite the mentioned limitations, in section 4.2, we compare each of the articles based on the two criteria of sensitivity and FP according to what the authors mentioned in their article. The main problem of the existing systems is the inability to accurately describe the object according to the structure of the suspected embolism mass. Future work could include examining the difference between emboli in the main branches and peripheral branches and dissecting the entire embolus.

**Conflict of Interest**

Dear Editor,

Hereby I certify that the authors declare that there is no conflict of interest regarding the publication of this article.

Best regards,
Seyed Hesamoddin Hosseini

M.Sc. of AI Engineer
Department of Computer Engineering, Ferdowsi University of Mashhad, Mashhad, Iran

**Data availability:**

Data sharing not applicable to this article as no datasets were generated or analyzed during the current study

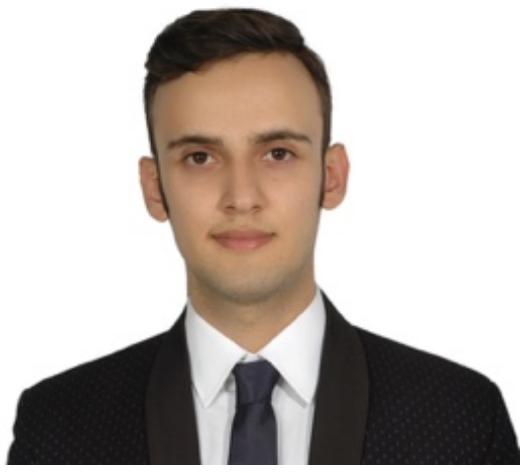

**Seyed Hesamoddin Hosseini** was born in Mashhad, Iran. He received the B.Sc. and M.Sc. degrees in Computer (Artificial Intelligence and Robotics) Engineering, from Islamic Azad University (IAU), Mashhad Branch, Mashhad, Iran (Jul.2021) and Ferdowsi University of Mashhad (FUM), Mashhad, Iran (Sep.2023), respectively. His research interests include the Medical Image processing, Biomedical Image Analysis, Machine Learning and Deep Learning.

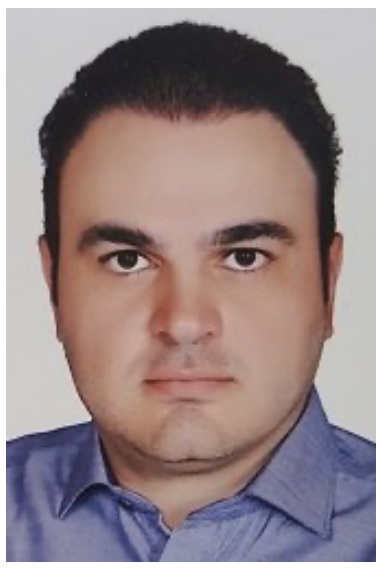

**Amir Hossein Taherinia** is an associate professor at the Computer Engineering Department, Ferdowsi University of Mashhad, Iran. He obtained his B.Sc in computer engineering from Ferdowsi University of Mashhad, Iran, in 2004. He obtained his M.Sc and PhD. degree in Artificial Intelligence from Sharif University of Technology, Tehran, Iran in 2006 and 2011, respectively. Afterwards, he started working as an assistant professor in Ferdowsi university of Mashhad, Mashhad, Iran. Where at 2020 he promoted to associate professor. His main research interests are Machine Learning and Machine Vision.

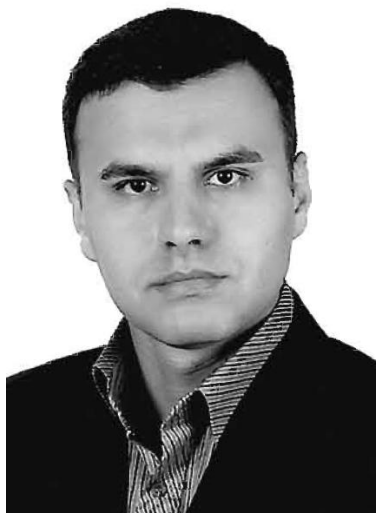

**Mehdi Saadatmand** was born in Iran, in 1980. He received the B.Sc. degree in Electrical (Control) Engineering from Ferdowsi University of Mashhad (FUM), Mashhad, Iran (2002), and the M.Sc. and Ph.D. degrees, both in Biomedical Engineering, from K.N. Toosi University of Technology, Tehran, Iran (2005) and Tarbiat Modares University, Tehran, Iran (2010), respectively. He is working as an Associate Professor in the Department of Electrical (Biomedical) Engineering of FUM. His fields of interest are medical imaging, medical image processing, computer-aided diagnosis, computer vision, and soft computing. He has published about 120 peer-reviewed articles in these research fields.